\documentclass[a4paper,10pt]{article}
\usepackage[utf8x]{inputenc}

\title{Super-Group Field Cosmology}
\author{ Mir Faizal \\
Mathematical Institute, University of Oxford
\\ Oxford
OX1 3LB, United Kingdom 
 }

\begin{document}

\maketitle

\begin{abstract} In this paper we construct a model for group field cosmology. 
The classical equations of motion for the non-interactive part of this model generate the Hamiltonian constraint 
of loop quantum gravity for a homogeneous isotropic  
universe filled  with a scalar matter field. The interactions  
represent topology changing processes that occurs due to joining and splitting of universes.   
These universes in the multiverse are assumed to obey both bosonic and fermionic statistics, and so a supersymmetric multiverse
is constructed using  superspace formalism. We  also introduce gauge symmetry  in this model. 
The supersymmetry and gauge symmetry are  introduced at the level of 
third quantized fields, and not the second quantized ones. This is the first time that  supersymmetry 
has been discussed at the level of third quantized fields. 
\end{abstract}
\section{Introduction}
Loop quantum gravity is an background independent way to quantize gravity, where the Hamiltonian constraint  is written in terms of 
Ashtekar-Barbero connection and densitized triad \cite{1}-\cite{1a}. 
In loop quantum gravity the curvature of the connection is 
expressed through 
the holonomy around a loop. The area of such a loop cannot take 
arbitrary small nonzero values because 
 the area operator in loop quantum gravity has a discrete spectrum.
 At a kinematics level 
the Hilbert space of loop quantum gravity is a space of spin networks 
\cite{s}-\cite{sa}, and the time evolution of these spin 
networks naturally 
leads to the spin foam model \cite{f}-\cite{fa}. 
 Even though in loop quantum gravity the geometry is dynamical, the topology is fixed. 
This is because, like the canonical formalism, loop quantum gravity is a second quantized formalism. 
Just like  the processes where the particle number is not conserved
cannot be fully analysed in the first quantization formalism, and we have to resort to second quantization to analyse such processes;
 the topology changing processes  cannot be fully analysed in second quantized formalism,
 and we have to resort to third quantization  to analyse such processes
 \cite{t}-\cite{ta}. 
A third quantization of loop quantum gravity naturally 
leads to group fields theory,
 which is a field theory defined on a group manifold \cite{g}-\cite{ga}.
 Topology changing process can be analysed in the framework of group field theory. These processes occur due to the creation and annihilation 
of spin states. So, the Feynman amplitudes of group field theory can be 
equivalently expressed in terms of spin-foam models. 
Many models for  loop quantum cosmology  have
 been recently analysed and are 
 better understood than full loop quantum gravity \cite{c}-\cite{ca}.
 Similarly, group field cosmology has been studied 
 as a mini-superspace approximation to the 
full group field theory \cite{m}-\cite{ma}. In group field cosmology first
the Hamiltonian constraint of loop quantum cosmology 
is viewed as the kinetic part of the group field theory and then  interactions are added. 
The bosonic group field cosmology has already been studied and 
 there is no reason to restrict the statistics of the group field cosmology to bosonic case. Hence, in this paper we will also analyse a fermionic 
group field cosmology. We will in fact construct a supersymmetric group field cosmology in ${\mathcal{N}}=1$ superspace formalism with gauge symmetry. 
Even though  color has already been 
incorporated in group field theory \cite{color}-\cite{colora}, it has not been so far been incorporated in any model of group field cosmology. 
Furthermore,  the supersymmetric loop quantum gravity have been studied only at the second quantized level \cite{super}-\cite{supera},
and  supersymmetry has never been studied at the level of third quantized loop quantum gravity or loop quantum cosmology.
  In fact, no third quantized 
theory, including group field theory, has ever been supersymmetrized. 
 Supersymmetry has been an important ingredient in the study 
of M-theory \cite{mtheory}-\cite{mtheory1} and many 
phenomenological models beyond the standard model are based 
on it \cite{standard}-\cite{standard1}. It also provides 
the an interesting 
candidate for dark matter \cite{dark}-\cite{dark1}.
 Hence, it will be interesting to incorporate it in group
 field cosmology at a 
third quantized level.  Motivated from this fact we also 
construct a supersymmetric group field theory 
with gauge symmetry in this paper.

\section{Loop Quantum Cosmology}
In loop quantum cosmology, the integration  is restricted to a fixed three dimensional cell of finite volume $ {\cal{V}}_0$.
 The integration is restricted in order to regulate the infinities coming from homogeneous fields
 being integrated over a non-compact space-time \cite{qazs}-\cite{qazr}. 
The main  variables in   loop quantum gravity are  Ashtekar-Barbero connection $A^i_a $ 
and $E^a_i $.  In the loop quantum gravity the holonomies of the connection are defined as operators. 
We will analyse the loop quantum cosmology for a massless scalar field $\phi$ in a spatially flat, homogeneous, and isotropic universe with
metric,
\begin{equation}
 ds^2 = -N^2(t) dt^2 + a^2 (t) \delta_{ab} dx^a dx^b.
\end{equation}
Here  $a(t)$ is the scalar factor and $N(t)$ is the lapse function. 
Now $A^i_a = \gamma (\omega^i_0)_a,  $ where $\gamma$ is the Barbero-Immirzi parameter and $(\omega^i_0)_a$ is the Levi-Civita connection. 
In loop quantum cosmology $A^i_a$  only depends on $c = \pm {\cal{V}} _0^{1/3} N^{-1} \gamma a'$,
 with $a'$ being the time derivative of $a$.  
Furthermore, the  variable promoted to operators are $p = \pm a^2 {\cal{V}} _0^{2/3}$,  and $exp(i\mu c)$, where $c$ is conjugate to $p$, and 
 $\mu$ is a function of $p$. 
In the improved dynamics scheme \cite{a}, $\mu = |p|^{-1/2}$, and this forms a basis of the eigenstates of the volume operator 
$\cal{V}$ with 
\begin{equation}
 {\cal{V}} |\nu \rangle = 2 \pi \gamma G |\nu| |\nu \rangle, 
\end{equation}
where $\nu = \pm a^2 {\cal{V}}_0 /2\pi \gamma G$ has the dimensions of length. 
The variable conjugate to $\nu$ is $b$ and so 
 the main variable for the gravitational sector  are 
$\nu$ and $\exp( i \lambda b)$, where $\lambda$ is a constant.
For the matter sector one choses the usual Schroedinger quantization scheme and 
the tensor product of the kinematic Hilbert space of the gravitational sector 
with the kinematic Hilbert space of the matter sector give the total Hilbert space of the theory. 

For the dynamics of the theory the curvature of $A^i_a$ is expressed through the holonomy around a loop. The area of such a loop cannot 
cannot be smaller than a fixed minimum area because 
the smallest  eigenvalue of the area operator in loop quantum gravity is nonzero.  
In Planck units the Hamiltonian constrain for a homogeneous isotropic universe with a massless scalar field $\phi$,  can thus be written as  
\begin{equation} 
 K ^2\Phi(\nu, \phi) = [E^2 - \partial^2_\phi] \Phi(\nu, \phi) =0,
\end{equation}
where we have  defined $ E^2 $ to be 
\begin{eqnarray}
 E^2 \Phi(\nu, \phi) &=& - [B(\nu)]^{-1}C^+(\nu) \Phi(\nu+4 , \phi)
 - [B(\nu)]^{-1}C^0(\nu)\Phi(\nu, \phi)  \nonumber \\ && - [B(\nu)]^{-1}C^-(\nu)\Phi(\nu-4, \phi).
\end{eqnarray}
Here we have set $\nu_0 =4$. 
For the usual choice of gauge in  loop quantum cosmology, we have 
\begin{equation}
 C^-(\nu) =C^+(\nu- 4 ),
\end{equation}
and so we need to only specify the functions  $C^+(\nu), C^0(\nu)$ and $B(\nu)$. 
These functions depend of the choice of the lapse function and the quantization scheme choose,
 for example with $N =1$  in improved dynamic scheme, we have \cite{a} 
\begin{eqnarray}
C^+(\nu)&=&\frac{1}{12\gamma\sqrt{2\sqrt{3}}}\left|\nu+2\right|\left| \left|\nu+1\right|-
\left|\nu+3\right| \right|,\nonumber 
\\C^0(\nu)&=&-C^+(\nu)-C^+(\nu-4),\nonumber
\\
B(\nu)&=&\frac{3\sqrt{2}}{8\sqrt{\sqrt{3}}\pi\gamma G}|\nu| \left| \left|\nu+1\right|^{\frac{1}{3}}
-\left|\nu-1\right|^{\frac{1}{3}} \right|^3.
\end{eqnarray}
 Similarly, in solvable loop quantum cosmology with $N =a^3$, we have \cite{as}
\begin{eqnarray}
C^+(\nu)=\frac{\sqrt{3}}{8\gamma}\left(\nu+2\right),&& C^0(\nu)=-\frac{\sqrt{3}}{4\gamma}\nu,
\nonumber\\
B(\nu)=\frac{1}{\nu}. &&
\end{eqnarray}
There is  also a super-selection for  subspace of the  form $| \nu +4n\rangle$, for all 
integers  $n$ and  some state $|\nu\rangle$. 
It may also be noted that the matter variable appear as the time variable in  loop quantum cosmology. 
\section{Group Field Cosmology}
Just like the wave function of first quantized theories is viewed as a classical field in second quantized formalism, the wave function of the 
second quantized theory should be viewed as a classical field in third quantized formalism. Hence, the wave function of loop quantum cosmology 
will now be viewed as the classical field of group field cosmology. 
So, we want a free bosonic field theory such that its classical equation of motion reproduce the Hamiltonian 
constraint for loop quantum gravity. 
The classical action for this bosonic field theory can be written as  \cite{m}
\begin{equation}
 S_{b} = \sum_\nu \int d\phi \, \,    \Phi (\nu, \phi)  K ^2 \Phi(\nu, \phi). 
\end{equation}
As $K^2$ is not diagonal in $\nu$, it is useful to expand $\Phi(\nu, \phi)$ as 
\begin{eqnarray}
 \Phi (b, \phi) &=& \sum_\nu \Phi(\nu, \phi)  exp ( i \nu b), \nonumber \\ 
\Phi ( \nu, \phi) &=& \frac{2}{\pi} \int_0^{\pi/2} db  \,\,  \Phi(b, \phi) exp (-i\nu b).
\end{eqnarray}
Now the bosonic action is given by 
\begin{equation}
 S_b = \frac{1}{\pi} \int d\phi \, \,  \int db  \,\,  \overline{\Phi}(b, \phi) \tilde K^2 \Phi(b, \phi),  
\end{equation}
where $\Phi(v, \phi) = \overline{\Phi} (\pi/2 -b , \phi)$ and 
the inverse of $ \tilde K^2$ is the propagator of the theory. 
To construct the Fock space, we expand these fields into modes which we promote to operators \cite{m}. 
Each of these will create or annihilate universes. 
The vacuum state is defined as the state annihilated 
by all annihilation  operator, 
$
 a_k |0\rangle =0,
$
and the Fock space is constructed as by the action of creation operators on this vacuum state. The vacuum state here corresponds to a state with no
geometry, matter field and topological structure. The topology, geometry and matter is created by the action of 
creation operators on this vacuum state. Interactions  will now correspond to an interaction of these universes. 
Hence, they will represent the process of splitting of or joining of universes.  Thus the topology changing 
processes can be studied by including interaction terms to this bosonic free field theory.
 Each interaction will also correspond to  a conservation of a different quantity during the process of interaction of these universes. 
So, for example the interaction potential of the form $\lambda \Phi^3( \nu, \phi)/3$, will correspond to a conservation of 
 both $b$ and $p_\phi$, because 
it will implement locality in both the scalar factor and the scalar field. 

There is no reason to assume that the statistics of the theory 
are bosonic and so we 
can also construct a fermionic model for group field cosmology. 
The choice of fermionic statics will change 
how distribution of different universes in the multiverse.
In fact, attempts to explain the cosmological constant
 from a bosonic distributions of universes in the multiverse
 have led to a wrong value for the cosmological constant
\cite{cos1}. It will be interesting to perform
 a similar analysis with a fermionic distribution of universes 
and try to calculate the value of the cosmological constant. 
 This is one of the motivations for studding a fermionic 
distribution of universes in the multiverse. 

Fermionic statics 
will not change the dynamics of a single universe because the fermionic 
theory  will also be required to generate 
the Hamiltonian constraint of loop quantum gravity. 
Thus, in analogy with a fermionic 
field in two dimensions \cite{susy},
  we define a fermionic field in group field cosmology as 
 $\Psi_a (\nu, \phi) = (\Psi_1(\nu, \phi), \Psi_2(\nu, \phi) )$.
The spinor indices are
 raised and lowered by the second-rank 
antisymmetric tensors $C^{ab}$ and $C_{ab}$, respectively, 
\begin{eqnarray}
 \Psi^a (\nu, \phi)&=&  C^{ab}\Psi_b(\nu, \phi), \nonumber \\ 
 \Psi_a(\nu, \phi) &=& \Psi^b (\nu, \phi)C_{ba}.
\end{eqnarray}
These second-rank 
antisymmetric tensors also satisfy $C_{ab}C^{cb} = \delta^c_a$.
So, we have $(\gamma^\mu)_{ab} = (\gamma^\mu)_a^c C_{cb} 
= (\gamma^\mu)_{ba}$ 
\cite{bsusy}. 
Now we  
define $K_\mu = (E, \partial_\phi) $ and $\eta^{\mu\nu} = (1, -1) $.
Here $K_\mu$ acts 
like $\partial_\mu$ of a ordinary two dimensional field  theory, 
so in analogy with   $\partial_{ab} = (\gamma^\mu \partial_\mu)_{ab}$ 
and $\partial^2 = \partial^{ab}\partial_{ab}/2$ \cite{susy},  
we also define  $ K_{ab} = (\gamma^\mu )_{ab}K_\mu $ and 
\begin{eqnarray}
 K^2 &=& \frac{1}{2} K^{ab} K_{ab} \nonumber \\ 
&=& \frac{1}{2} (\gamma^\mu )^{ab}K_\mu(\gamma^\nu )_{ab}K_\nu
 \nonumber \\ 
&=& \eta^{\mu \nu } K_\mu K_\mu 
 \nonumber \\ 
&=& [E^2 - \partial^2_\phi],   
\end{eqnarray}
where we have used $\{\gamma^\mu, \gamma^\nu\} = 2 \eta^{\mu\nu}$. 
Now we can write  the action for the fermionic group field cosmology
 in analogy 
with the regular fermonic action in two dimensional 
spinor formalism 
by using the correspondence between 
$K_\mu$ and $\partial_\mu$ \cite{susy}, 
\begin{eqnarray}
 S_{f} &=& \sum_\nu \int d\phi \, \, C^{bc} \Psi_c (\nu, \phi)
 (\gamma^\mu )_{b}^a K_\mu \Psi_a(\nu, \phi)\nonumber \\ 
&=& \sum_\nu \int d\phi \, \,  \Psi^b(\nu, \phi) K_b^a \Psi_a(\nu, \phi).
\end{eqnarray}
The classical equations of motion obtained from this action are 
\begin{equation}
 K^b_a \Psi_b (\nu, \phi) =0. 
\end{equation}
Now action on it by $K^c_b$, and 
using $\{\gamma^\mu, \gamma^\nu\} = 2 \eta^{\mu\nu}$, we  get 
\begin{equation}
 K^2 \Psi_c (\nu, \phi) =0. 
\end{equation}
We can also construct the Fock space for this fermionic theory by first  expand these fields into modes and then promoting those modes  to operators. 
Just like the bosonic theory, the vacuum state of the fermionic theory will also be annihilated by all annihilation operators 
$
 a_k |0\rangle =0,
$
and the Fock space will be constructed  by the action 
of creation operators on this vacuum state. 
 This vacuum state again corresponds to a state with no
geometry, matter field and  topological structure.
Here the again modes denoted by $k$ correspond 
to the  momentum $p_\phi$, which is the momentum 
associated with the scalar field $\phi$ \cite{m}. 
We can also introduce interactions for this fermionic theory or 
an combination of both the
fermionic and bosonic theories. 
\section{Supersymmetry} 
In this section we will derive a supersymmetric generalization of the group field cosmology. 
We will in fact construct a  super-group field  theory, such that the kinetic parts of its  bosonic  and fermionic component fields will
generate the Hamiltonian constraint of loop quantum gravity. 
 In order to
do that  we first introduce $\theta_a$ as
two component anti-commuting parameters with odd Grassmann parity and let 
$
 \theta^2 =  \theta_a C^{ab} \theta_b/2 =  \theta^a \theta_a/2.
$
Now  using this Grassman variable $\theta_a$, we define the generators of ${\cal_{N}} =1$ supersymmetry as 
\begin{equation}
 Q_a = \partial_a - K_a^b \theta_b.
\end{equation}
The covariant derivative that commutes with this generator of supersymmetry can be written as 
\begin{equation}
 D_a = \partial_a + K^b_a \theta_b,
\end{equation}
where 
\begin{equation}
 \partial_a = \frac{\partial}{\partial \theta^a }.
\end{equation}
Hence, the full supersymmetric algebra between $D_a$ and $Q_a$ is given by 
\begin{eqnarray}
 \{Q_a, Q_b\} &=& (\partial_a - K_a^c \theta_c) (\partial_b - K_b^d \theta_d)  \nonumber \\ 
 &&  + (\partial_b - K_b^d \theta_d) (\partial_a - K_a^c \theta_c)\nonumber \\ &=&  2 K_{ab}, \nonumber \\ 
\{ D_a, D_b\} &=& (\partial_a + K_a^c \theta_c) (\partial_b + K_b^d \theta_d)  \nonumber \\ 
 &&  + (\partial_b + K_b^d \theta_d) (\partial_a + K_a^c \theta_c)\nonumber \\ &=& - 2 K_{ab}, \nonumber \\
\{Q_a, D_b\} &=& (\partial_a - K_a^c \theta_c) (\partial_b + K_b^d \theta_d)  \nonumber \\ 
 &&  + (\partial_b +  K_b^d \theta_d) (\partial_a - K_a^c \theta_c)\nonumber \\ &=&  0,
\end{eqnarray}
because $(\gamma^\mu K_\mu)_{ab} = (\gamma^\mu K_\mu)_a^c C_{cb} = (\gamma^\mu K_\mu)_{ba}$ \cite{bsusy}.
Now we will write a supersymmetric theory in $ {\mathcal{N}}= 1$ superspace formalism so that it has manifest  $ {\mathcal{N}}= 1$ supersymmetry. 
To do that, we define a superfield $\Omega( \nu, \phi, \theta)$, such that 
\begin{equation}
 \Omega( \nu, \phi, \theta) = \Phi( \nu, \phi) + \theta^a \Psi_a( \nu, \phi) - \theta^2 F ( \nu, \phi).
\end{equation}
 Now we can write
\begin{eqnarray}
\Phi( \nu, \phi) = [\Omega( \nu, \phi, \theta)]_|, && \Psi_a( \nu, \phi)=  [D_a \Omega( \nu, \phi, \theta)]_|, \nonumber \\ 
F ( \nu, \phi)= [D^2 \Omega( \nu, \phi, \theta)]_|, 
\end{eqnarray}
where $'|'$ means we set $\theta_a =0$ at the end of calculation.  
The supersymmetric transformations $ \delta_{s}$ generated by $Q_a$ can also be calculated in analogy with the regular supersymmetric theory, 
\begin{eqnarray}
 \delta_{s} \Phi( \nu, \phi) &=& - \epsilon^a ( \nu, \phi) \Psi_a ( \nu, \phi), \nonumber \\ 
\delta_{s} \Psi_a ( \nu, \phi) &=&- \epsilon^b( \nu, \phi) [C_{ab} F ( \nu, \phi)  + K_{ab}\Phi( \nu, \phi) ], \nonumber \\ 
\delta_{s} F ( \nu, \phi)&=&-\epsilon^a ( \nu, \phi)  K_a^b \Psi_b ( \nu, \phi).
\end{eqnarray}
These transformations are similar to the regular $\mathcal{N} =1$ transformations in three dimensions \cite{susy}, 
with the spacetime replaced by $(\nu, \phi)$ and $\partial_{ab}$ replaced by $K_{ab}$. 
Now we can write a superfield theory in 
this  $\mathcal{N} =1$  superspace. As the field theory will be written in $\mathcal{N} =1$, it will have manifest supersymmetry. 
However, we have two operators $E$ and
$\partial_\phi$, so we have to work in analogy with a two dimensional superfield theory. It is well known that 
$\mathcal{N} =1$ supersymmetry in three dimensions corresponds to $\mathcal{N} =2$ supersymmetry in two dimensions. This is because 
by using the projection  $P_{\pm } = (1 \pm \gamma^3)/2$, 
we can split the super-charge $Q_a$  as 
$\epsilon^a Q_a = \epsilon^- ( \nu, \phi) Q_- + \epsilon^- ( \nu, \phi) Q_+$. 
In two dimensions these $Q_+$ and $Q_-$ act as two independent super-charges \cite{bsusy}-\cite{susy1}. So, 
our theory actually has $\mathcal{N} =2$ supersymmetry.
Thus, again in analogy with regular free supersymmetric theory, we write the following action, 
\begin{eqnarray}
 S_{s} &=& \frac{1}{2}\sum_\nu \int d\phi \, \, 
 [D^2 [\Omega ( \nu, \phi, \theta) D^2 \Omega ( \nu, \phi, \theta)]]_| 
\nonumber \\
&=& \frac{1}{2} \sum_\nu \int d\phi \, \, 
[ D^2 \Omega( \nu, \phi, \theta) D^2 \Omega( \nu, \phi, \theta)
 \nonumber \\ &&
 \,\,\,\,\,\,\,\,\,\,\,\,\,\,\,\,\,\,\,\,\,\,\,+ D^a\Omega( \nu, \phi, \theta) D_a D^2 \Omega( \nu, \phi, \theta) \nonumber \\ &&
 \,\,\,\,\,\,\,\,\,\,\,\,\,\,\,\,\,\,\,\,\,\,\,+ \Omega ( \nu, \phi, \theta) (D^2)^2 \Omega( \nu, \phi, \theta)]]_|
\nonumber \\ &=& \frac{1}{2}
 \sum_\nu \int d\phi \, \,
  [ F^2( \nu, \phi) + \Psi^b(\nu, \phi)
 K_b^a \Psi_a(\nu, \phi) \nonumber \\ &&
 \,\,\,\,\,\,\,\,\,\,\,\,\,\,\,\,\,\,\,\,\,\,\,
+  \Phi (\nu, \phi)  K ^2 \Phi(\nu, \phi) ].
\end{eqnarray}
This is the required supersymmetric model whose bosonic part and fermionic parts generated the bosonic and fermionic 
models of loop quantum cosmology. 
We can now add interactions to this supersymmetric theory,  
\begin{eqnarray}
  S_{si} &=& \sum_\nu \int d\phi \, \,  [D^2  [ f( \Omega( \nu, \phi, \theta))]]_| \nonumber \\ 
&=& \sum_\nu \int d\phi \, \,  [ f'' ( \Omega( \nu, \phi, \theta)) (D^a \Omega( \nu, \phi, \theta))^2 \nonumber \\ &&
 \,\,\,\,\,\,\,\,\,\,\,\,\,\,\,\,\,\,\,\,\,\,+ f'(\Omega( \nu, \phi, \theta)) D^2 \Omega( \nu, \phi, \theta)]_|   \nonumber \\ 
&=&    \sum_\nu \int d\phi \, \,  [f''(\Phi)( \nu, \phi) \Psi^2 ( \nu, \phi) + f'(\Phi)( \nu, \phi) F( \nu, \phi) ],
\end{eqnarray}
where $f( \Omega( \nu, \phi, \theta))$ is a general interaction term containing polynomial terms in $\Omega( \nu, \phi, \theta)$
 and $\Psi^2 ( \nu, \phi)= \Psi^a( \nu, \phi) \Psi_a( \nu, \phi)/2 $. 
For example, the  superspace interaction term given by 
\begin{equation}
f( \Omega( \nu, \phi, \theta)) = \frac{\lambda}{6} \Omega^3( \nu, \phi, \theta),  
\end{equation}
generates  the following component action, 
\begin{eqnarray}
 S_t &=& \frac{1}{2}\sum_\nu \int d\phi \, \,  
[ F^2( \nu, \phi) +  \Psi^b(\nu, \phi) K_b^a \Psi_a(\nu, \phi) 
\nonumber \\ &&
 \,\,\,\,\,\,\,\, \,\,\,\,\,\,\,\,\,\,\,\,\,\,\,\,\,\,\,\,\,
+  \Phi (\nu, \phi)  K ^2 \Phi(\nu, \phi)  
+ 2\lambda \Phi (\nu, \phi)\Psi^2(\nu, \phi)  \nonumber \\ &&
  \,\,\,\,\,\,\,\,\,\,\,\,\,\,\,\,\,\,\,\,\,\,\,\,\,\,\,\,\, 
+ \lambda \Phi^2(\nu, \phi) F(\nu, \phi) ],
\end{eqnarray}
where $S_t =  S_{s} + S_{si}$.
It may be noted that the exact nature of the group field cosmology will effect the way individual universes interact in the multiverse. 
As it is not possible at present to view other universes, it seems difficult to see which of these models is close
 to being the correct low energy description to some full quantum theory of gravity that describes the full multiverse. One way to deal with, 
this problem is to view these models as describing the quantum states of our universe at Planck level and then trying to analyse it to see if 
it can lead to some effective phenomenological description, at energy scales that we can measure.  
So, it is  interesting consider these as patches  of the our universe that interact with one another.
Such a possibility has been discussed and used to analyse cosmological perturbations \cite{pat}-\cite{pat2}.  
This possibility  has also been discussed in the framework of bosonic group field cosmology \cite{m}. 
\section{Gauge Symmetry}
In analogy with regular particle physics, we can also incorporate gauge  symmetry into the 
 third quantized group field cosmology. In fact, in this section, 
 we 
will constraint  supersymmetric group field cosmology  with gauge symmetry. To do that we first consider complex superfields 
$\Omega(\nu, \phi, \theta) $ and $ \Omega^{\dagger}  (\nu, \phi, \theta)$, which are suitably contracted with generators of a Lie algebra,
  $\Omega(\nu, \phi, \theta) = \Omega^A(\nu, \phi, \theta) T_A,$ and $ 
\Omega^{\dagger}(\nu, \phi, \theta) = \Omega^{\dagger A}(\nu, \phi, \theta)  T_A$. Here  
 $T_A$ are Hermitian generators of a Lie algebra 
$[T_A, T_B] = i f_{AB}^C T_C$. 
Now we define 
\begin{eqnarray}
 \Omega^A(\nu, \phi, \theta) &=& \Phi^A( \nu, \phi) + \theta^a \Psi^A_a( \nu, \phi) - \theta^2 F^A ( \nu, \phi), \nonumber \\ 
 \Omega^{A \dagger}(\nu, \phi, \theta) &=& \Phi^{A \dagger}( \nu, \phi) + \theta^a \Psi^{A \dagger}_a( \nu, \phi)
 - \theta^2 F^{A \dagger} ( \nu, \phi),\nonumber \\ 
\Lambda^A (\nu, \phi, \theta) &=& \lambda^A (\nu, \phi)  + \theta^a  \mu_a^A (\nu, \phi) - \theta^2 \nu^A ( \nu, \phi).
\end{eqnarray}
These superfields transform under infinitesimal gauge
 transformations as
\begin{eqnarray}
  \delta \Omega^A(\nu, \phi, \theta) &=&  if_{CB}^A\Lambda^C (\nu, \phi, \theta)\Omega^B(\nu, \phi, \theta) ,\nonumber\\
\delta \Omega^{A \dagger}(\nu, \phi, \theta)  &=& -i f^A_{CB}\Omega^{C\dagger}(\nu, \phi, \theta) \Lambda^B(\nu, \phi, \theta).
\end{eqnarray}
These transformations can now be written as 
\begin{eqnarray}
 \delta \Phi^A( \nu, \phi) &=&  i f_{CB}^A\lambda^C (\nu, \phi) \Phi^B( \nu, \phi), \nonumber \\ 
 \delta \Phi^{A \dagger}( \nu, \phi) &=&  - i f_{CB}^A (\nu, \phi) \Phi^{C\dagger}( \nu, \phi)\lambda^{B\dagger}, \nonumber \\ 
\delta\Psi^{A }_a( \nu, \phi) &=& if_{CB}^A [\lambda^C (\nu, \phi) \Psi^B_a( \nu, \phi) +\mu_a^C (\nu, \phi) \Phi^B( \nu, \phi) ],  \nonumber \\ 
\delta\Psi^{A \dagger}_a( \nu, \phi) &=& - if_{CB}^A [  \Psi^{C \dagger}_a( \nu, \phi)\lambda^{B \dagger}(\nu, \phi)
 + \Phi^{C \dagger}( \nu, \phi) \mu_a^{B \dagger} (\nu, \phi)],\nonumber \\ 
\delta F^A ( \nu, \phi) &=&  i f_{CB}^A [\lambda^C (\nu, \phi) F^B ( \nu, \phi) + \nu^C ( \nu, \phi)\Phi^B( \nu, \phi)
 \nonumber \\  && \,\,\,\,\,\,\,\,\,\,\,\, + \mu^{C a} (\nu, \phi)\Psi^B_a( \nu, \phi)], \nonumber \\ 
\delta F^{A \dagger} ( \nu, \phi) &=& - i f_{CB}^A [ F^{C\dagger} ( \nu, \phi) \lambda^{B\dagger} (\nu, \phi)
 + \Phi^{C\dagger} ( \nu, \phi)\nu^{B\dagger} ( \nu, \phi)
 \nonumber \\  && \,\,\,\,\,\,\,\,\,\,\,\,+\Psi^{C a\dagger}( \nu, \phi) \mu_a^{B\dagger} (\nu, \phi)].
\end{eqnarray}
 Now the  super-derivative,  given by 
$
 D_a = \partial_a + K^b_a \theta_b,
$
of these superfields does not transform 
 like the original superfields. But 
 we can define a super-covariant derivative for  these superfields 
by requiring it   to transform like the original superfields. 
 Thus, we obtain the following expression for the     
 super-covariant derivative of these superfields
\begin{eqnarray}
  \nabla_a  \Omega^A(\nu, \phi, \theta)&=& D_a\Omega^A(\nu, \phi, \theta) -i f_{CB}^A\Gamma^C_a(\nu, \phi, \theta) \Omega^B(\nu, \phi, \theta),\nonumber\\
\nabla_a \Omega^{A \dagger}(\nu, \phi, \theta)  &=& D_a \Omega^{A \dagger}(\nu, \phi, \theta)  
+ i  f_{CB}^A\Omega^{C \dagger}(\nu, \phi, \theta)  \Gamma^B_a (\nu, \phi, \theta).
\end{eqnarray}
If  the spinor superfield $\Gamma_a^A (\nu, \phi, \theta) $ is made to transform under 
gauge transformations as 
\begin{equation}
 \delta \Gamma^A_a(\nu, \phi, \theta)  = \nabla_a \Lambda^A(\nu, \phi, \theta),
\end{equation}
then the  super-covariant derivative of the
 scalar superfields $\Omega^A(\nu, \phi, \theta)$ and $\Omega^{A \dagger}(\nu, \phi, \theta)$ transforms
 under gauge   transformations like the original fields,
\begin{eqnarray}
  \delta \nabla_a \Omega^{A \dagger}(\nu, \phi, \theta)&=&  if_{CB}^A\Lambda^C (\nu, \phi, \theta) \nabla_a\Omega^{B \dagger}(\nu, \phi, \theta) ,\nonumber\\
\delta \nabla_a\Omega^{A \dagger}(\nu, \phi, \theta) &=& -i f_{CB}^A \nabla_a\Omega^{C \dagger}(\nu, \phi, \theta)\Lambda^B(\nu, \phi, \theta).
\end{eqnarray}

Now we can use these superfields to write an action for  supersymmetric group field cosmology with gauge symmetry.
To that we first define $\Gamma_a(\nu, \phi, \theta)$ to be a matrix valued spinor
 superfield which is suitable contracted with generators of a Lie algebra,  $ \Gamma_a(\nu, \phi, \theta) = \Gamma_a^A(\nu, \phi, \theta) T_A $.
Now define the components of this superfield  $\Gamma_a(\nu, \phi, \theta)$ to be
\begin{eqnarray}
A_a (\nu, \phi)&= &[\Gamma_a(\nu, \phi, \theta)]_|, \nonumber \\ 
 X(\nu, \phi)&= & -\frac{1}{2}[D^a \Gamma_a(\nu, \phi, \theta)]_|, \nonumber \\ 
A^\mu (\nu, \phi)&= &- \frac{1 }{2} [ D^a (\gamma^{\mu })_a^b \Gamma_b (\nu, \phi, \theta)]_|, \nonumber \\
 X_a(\nu, \phi) &= &\frac{1 }{2}[D^b D_a  \Gamma_b(\nu, \phi, \theta)]_|. \label{csf}
\end{eqnarray}
It may be noted that the  gauge field $A^\mu (\nu, \phi)$ in group field  cosmology is analogy to ordinary gauge field. 
Thus, $\Gamma_a(\nu, \phi, \theta)$ is the supersymmetric generalization of the  gauge field in group field cosmology. 
We also 
 define a field strength for this supersymmetric gauge field in group field cosmology as, 
\begin{equation}
 \omega_a (\nu, \phi, \theta) = \nabla^b \nabla_a \Gamma_b(\nu, \phi, \theta). 
\end{equation}
Now we can write the action for the gauge theory as 
\begin{eqnarray}
S_{ga} &= &\sum_\nu \int d\phi \, \,    [D^2 [\Omega^{\dagger} ( \nu, \phi, \theta) \nabla^2 \Omega ( \nu, \phi, \theta) 
\nonumber \\ && \,\,\,\,\,\,\,\,\,\,\,\,\,\,\,\,\,\,\,\,\,\,\,\,\, + \omega^a ( \nu, \phi, \theta) \omega_a( \nu, \phi, \theta)  ]]_|. 
\end{eqnarray}
As this theory has a gauge symmetry, we need to fix a gauge before we can quantize it. Thus, in analogy with 
ordinary supersymmetric gauge theory \cite{susy}, we chose the following gauge, 
\begin{equation}
 D^a \Gamma_a ( \nu, \phi, \theta) =0.
\end{equation}
This can be incorporated at a quantum level by adding the following gauge fixing term to original classical action, 
\begin{equation}
 S_{gf} = \sum_\nu \int d\phi \, \,    [D^2[ B ( \nu, \phi, \theta) D^a \Gamma_a ( \nu, \phi, \theta)]]_|, 
\end{equation}
where $B( \nu, \phi, \theta) = B^A( \nu, \phi, \theta) T_A$ is a 
auxiliary superfield. This field can be integrated out, in the path 
integral formalism, to impose the 
gauge fixing condition. Now we can write the ghost term by  gauge transforming the gauge fixing condition, replacing 
$\Lambda(\nu, \phi, \theta) = \Lambda^A (\nu, \phi, \theta) T_A$
 by ghost superfield $ {C}( \nu, \phi, \theta) =  {C}^A( \nu, \phi, \theta)  T_A$
and then  contracted it with anti-ghost superfield $ \overline{C}( \nu, \phi, \theta) = \overline{C}^A( \nu, \phi, \theta)  T_A$ \cite{fdfd},   
\begin{equation}
  S_{gh} = \sum_\nu \int d\phi \, \,   
 [D^2[\overline C( \nu, \phi, \theta) D^a \nabla_a C 
( \nu, \phi, \theta)]]_|.
\end{equation}
Thus, the sum of the  ghost and gauge fixing terms is given by 
\begin{eqnarray}
 S_{gh} + S_{gf} &=& \sum_\nu \int d\phi \, \,    [D^2[ B ( \nu, \phi, \theta) D^a \Gamma_a ( \nu, \phi, \theta) + \nonumber \\ 
&& \,\,\,\,\,\,\,\,\,\,\,\,\,\,\,\,\,\,\,\,\,\,\,\,\,
\overline C( \nu, \phi, \theta) D^a \nabla_a C ( \nu, \phi, \theta)]]_|.
\end{eqnarray}
It may be noted that in analogy with ordinary supersymmetric gauge theory the a mixing of $D_a$ and $\nabla_a$ 
occurs in the ghost term. 
 Now  we can write the vacuum to vacuum transition as 
\begin{equation}
 Z = \int D M \,  exp (iS), 
\end{equation}
where $DM =  D\Omega^{\dagger } D \Omega DC D\overline{C} DB D\Gamma_a $ and 
$ S = S_{ga}+  S_{gh}+  S_{gf}$, without setting $\theta_a = 0$. Now we can derive the propagators for this theory and the Feynman
 rules using standard field theory techniques. 
What is
interesting is that this action allows for  processes like  a bosonic universe with the wave function $\Phi( \nu, \phi) $ or a 
fermionic universe with the wave function $\Psi(\nu, \phi)$ to be interact with each other via exchange 
of third quantized  virtual gauge universe $A_\mu (\nu, \phi)$. Thus, the big bang in this model can be viewed as creation of a pair of universe 
by the interaction of two other universes via the exchange of some third quantized virtual gauge universe. 
Just like the first quantized wave function is 
viewed as a classical field in second quantization, the second quantized wave function can be viewed as a classical 
field in third quantization. Now, we know that the particle creation can only be consistently explained 
in the framework of a second quantized theory. From the perspective of a first quantized theory the wave function of 
a particle will suddenly disappear when it get annihilated. Similarly, the wave function of a particle will suddenly appear 
when it forms. However, from the perspective of a second quantized theory when the first quantized wave function is 
viewed as a classical field, the disappearance or the appearance of a particle can be easily understood in the light of an 
interactive field theory. Similarly,  from the perspective of second quantization the wave function of the universe suddenly appears at the big bang. 
Similarly, the wave function of a black holes seems to disappear during the process of evaporation of a black hole. However, 
from a third quantized perspective, it is just the creation and annihilation of these spacetimes and matter configurations, that can 
be understood in the light of interactive group field cosmology.

In order to consistently analyse the supersymmetric gauge field cosmology we need to analyse the third quantized 
BRST symmetry of the theory. These symmetries can be use to show that various topology changing 
processes are unitarity from a third quantized perspective.  So,  we will analyse the third quantized BRST  symmetry for this theory. 
The sum of the ghost term with the  gauge fixing term can  be written as a total third quantized BRST variation, 
\begin{equation}
  S_{gh} + S_{gf} = \sum_\nu \int d\phi \, \,    s \, [D^2 [\overline{C}  ( \nu, \phi, \theta) D^a \Gamma_a ( \nu, \phi, \theta) ]]_|,
\end{equation}
where the third quantized BRST transformations are given by 
\begin{eqnarray}
  s\,\Omega^A(\nu, \phi, \theta)&= &if_{CB}
  ^AC^C (\nu, \phi, \theta)\Omega^B(\nu, \phi, \theta) , \nonumber \\
s\,  \Omega^{A \dagger}(\nu, \phi, \theta)  &= & -i f^A_{CB}\Omega^{C\dagger}(\nu, \phi, \theta) C^B(\nu, \phi, \theta), 
\nonumber \\ 
s\, C^A( \nu, \phi, \theta)&= &f^A_{CB}C^{C\dagger}(\nu, \phi, \theta) C^B(\nu, \phi, \theta), 
\nonumber \\
 s\, \Gamma^A_a ( \nu, \phi, \theta) &= & \nabla_a C^A( \nu, \phi, \theta), 
\nonumber \\ 
s\, \overline{C} ^A( \nu, \phi, \theta) &= & B^A( \nu, \phi, \theta),
 \nonumber \\
 s\, B^A( \nu, \phi, \theta) &= &0.
\end{eqnarray}
Here we have  obtained third quantized BRST by following the 
procedure for obtaining the   BRST transformations of any gauge field theory \cite{BRST}-\cite{BRST1}. 
 Thus, the third quantized BRST transformations of the gauge and matter fields is there gauge transformations 
with the gauge parameter replaced by the ghost superfield. The third quantized BRST transformation of the 
ghost superfield is given by by contacting two ghost
superfields with the structure constant of the gauge group and the third quantized BRST transformation of anti-ghosts is 
the auxiliary superfield used to impose
gauge fixing condition. Finally, The third quantized BRST transformation of the auxiliary field vanishes. 
 These third quantized BRST 
transformations are nilpotent, $s^2 =0$, and so the sum of the ghost term with the gauge fixing term is invariant under these third quantized 
BRST transformations.
As the third quantized BRST transformations of the original third quantized classical action is only a ghost valued gauge transformations, 
it is also invariant
under these third quantized BRST transformations. Thus, the full action is invariant under these transformations, $
 s\, S  =0$. These third quantized BRST transformation can be used to construct a third quantized BRST charge and analyse
 the unitarity of the supersymmetric group field 
cosmology with gauge symmetry.

\section{Conclusion}
In this paper we have analysed the group field cosmology.
 To do so, we have analysed the loop quantum cosmology in third quantized formalism. Thus, after reviewing the 
basics of loop quantum cosmology, we constructed both 
 bosonic and fermionic field theories   whose classical equations of motion  generated the 
the Hamiltonian constraint of loop quantum cosmology. We  discussed the effect of interactions for these models of group field cosmology. 
We also constructed a supersymmetric theory in $ {\mathcal{N}} = 1 $ superspace 
formalism with gauge symmetry. 
 The component fields of this supersymmetric gauge theory  satisfied the Hamiltonian constraint of loop quantum 
cosmology.
We  analysed the interactions in this supersymmetric  theory and derived the corresponding interactions for the bosonic and 
fermionic fields component fields theories. In the supersymmetric gauge theory, two real universes interacted by an exchange of virtual universes. 
The big bang in this model was  viewed as creation of a pair of universe 
by the interaction of two other universes via the exchange of some third quantized virtual gauge universe. 
We also analysed the third quantized BRST of this supersymmetric group field theory with gauge symmetry. 
 We stress the fact that we did not 
supersymmetrize the theory at the level of second quantized fields but at 
the level of third quantized fields. 
This is the first time that supersymmetry has been studied at the level of a third quantized field theory.  

It would be interesting to carry to analyse further. One direction to carry out this analyses would be to perform it for 
the full group field theory. Both the fermionic and bosonic group field theory have been studied and color has also been 
incorporated in group field theory \cite{color}-\cite{colora}. It would thus be interesting to
analyse a supersymmetric group field theory with gauge symmetry.
 We will be able to develop that theory in analogy with the usual gauge theory. 
It will be possible to derive this  theory in superspace formalism with manifest supersymmetry. 
 This analysis can also be initially done in the ${\mathcal{N}}= 1$ 
superspace formalism. however, it will  be nice to analyse this present theory or the full group theory in the 
extended superspace formalism, with higher amount of manifest supersymmetry. 
It would also be interesting to analyse string field theory via group field cosmology.
 String theory can be viewed as two dimensional gravity 
coupled to matter fields. Thus, it can be quantized via loop 
quantum gravity. After that a group field cosmology of string theory 
can be constructed. This 
way we will be able to incorporate higher order string interactions in 
string field theory. In fact, the   Hamiltonian constraint for 
string in loop quantum gravity has already been analysed 
\cite{stringfi}. 
Now this 
 Hamiltonian constraint string in loop quantum gravity,
$K^2_{string}\Phi =0$,  can view this as a kinetic part 
of a group field theory action, $\Phi K^2_{string} \Phi$.
Then interactions can and introduce  through polynomial 
interaction $\Phi^n$. 
Hence, we will be able to go beyond the regular 
cubic string field theory by using group field cosmology \cite{stringfi1}. 
In fact, we can use all the results of this paper to analyse 
this string field theory. 
We could also try to construct the a superstring field theory of  
 heterotic string as a group field 
theory with the symmetry group $E_8\times E_8$ or $Spin(32)/Z_2$. 
We can then incorporate arbitrary interactions for these string field 
theories
via the standard methods of group field theory.

\end{document}